\documentclass[reprint,aps,prl,superscriptaddress]{revtex4-2}

\usepackage{graphicx}
\graphicspath{{./figs/}}

\usepackage{dcolumn}
\usepackage{bm}
\usepackage{hyperref}
\usepackage[svgnames]{xcolor}
\usepackage[mathlines]{lineno}
\usepackage{miller}

\usepackage{siunitx}
\DeclareSIUnit{\dbm}{dBm}

\usepackage{here}
\usepackage{amsmath}
\usepackage{bm}
\usepackage{graphicx}
\usepackage{mathrsfs}

\begin{document}
\title{Observation of spin-splitter torque in collinear antiferromagnetic RuO$_2$}
\author{Shutaro Karube}
\email{karube@material.tohoku.ac.jp}
\affiliation{Department of Materials Science, Tohoku University, Sendai 980-8579, Japan}
\affiliation{Center for Spintronics Research Network, Sendai 980-8577, Japan}

\author{Takahiro Tanaka}
\affiliation{Department of Materials Science, Tohoku University, Sendai 980-8579, Japan}

\author{Daichi Sugawara}
\affiliation{Department of Materials Science, Tohoku University, Sendai 980-8579, Japan}

\author{Naohiro Kadoguchi}
\affiliation{Department of Materials Science, Tohoku University, Sendai 980-8579, Japan}

\author{Makoto Kohda}
\affiliation{Department of Materials Science, Tohoku University, Sendai 980-8579, Japan}
\affiliation{Center for Spintronics Research Network, Sendai 980-8577, Japan}
\affiliation{Center for Science and Innovation in Spintronics (Core Research Cluster), Sendai 980-8577, Japan}
\affiliation{Division for the Establishment of Frontier Sciences of the Organization for Advanced Studies, Sendai 980-8577, Japan}

\author{Junsaku Nitta}
\affiliation{Department of Materials Science, Tohoku University, Sendai 980-8579, Japan}
\affiliation{Center for Spintronics Research Network, Sendai 980-8577, Japan}
\affiliation{Center for Science and Innovation in Spintronics (Core Research Cluster), Sendai 980-8577, Japan}
 
\begin{abstract}
    \quad The spin-splitter effect is theoretically predicted to generate an unconventional spin current with $\mathit{x}$- and $\mathit{z}$- spin polarization via the spin-split band in antiferromagnets. The generated torque, namely spin-splitter torque, is effective for the manipulation of magnetization in an adjacent magnetic layer without an external magnetic field for spintronic devices such as MRAM. Here, we study the generation of torque in collinear antiferromagnetic RuO$_2$ with (100), (101), and (001) crystal planes. Next we find all $\mathit{x}$-, $\mathit{y}$-, and $\mathit{z}$-polarized spin currents depending on the N\'{e}el vector direction in RuO$_2$(101). For RuO$_2$(100) and (001), only $\mathit{y}$-polarized spin current was present, which is independent of the N\'{e}el vector. Using the $\mathit{z}$-polarized spin currents, we demonstrate field-free switching of the perpendicular magnetized ferromagnet at room temperature. The spin-splitter torque generated from RuO$_2$ is verified to be useful for the switching phenomenon and paves the way for a further understanding of the detailed mechanism of the spin-splitter effect and for developing antiferromagnetic spin-orbitronics. 
\end{abstract} 
\date{\today}
\maketitle

Remarkable breakthroughs in antiferromagnets (AFMs)~\cite{Baltz_RevModPhys_2018, Manchon_RevModPhys_2019} have been achieved in spintronics regarding exchange bias-induced field-free switching~\cite{Fukami_NatMater_2016, Park_NatNanotechnol_2016, Lau_NatNanotechnol_2016}, magnetic spin Hall effect (MSHE)~\cite{Kimata_Nature_2019, Hoffmann_PRL_2020}, and antiferromagnetic spin Hall effect (AFM-SHE)~\cite{Yang_NatMater_2021}. These intriguing phenomena have led to the establishment of antiferromagnetic spin-orbitronics to further manipulate magnetization effectively for spintronic devices because they are free from the restriction of symmetry regarding spin polarization of the generated spin current, like the conventional spin Hall effect (SHE). While the somewhat novel phenomena concerning unconventional spin-orbit torque in AFMs, such as IrMn$_3$~\cite{Hoffmann_PRL_2020}, PtMn$_3$~\cite{Song_PRB_2021}, and Mn$_{2}$Au~\cite{Yang_NatMater_2021} using 5d elements, still depend on spin-orbit coupling, it is difficult to uncover the mechanism generating the torque via the magnetic order. Recently, a spin-split band, through the antiferromagnetic order, was theoretically predicted to cause non-trivial torque, such as $x$- and $z$-polarized damping-like (DL) torque, even without spin-orbit coupling. This is referred to as the spin-splitter effect (SSE)~\cite{Jungwirth_PRL_2021}. Although there is theoretical evidence for the SSE, it has not yet been demonstrated. It is worth noting that there are differences in spin current generation among conventional SHE~\cite{Dyakonov_JETP_1971, Kimura_PRL_2007, Saitoh_APL_2006}, AFM-SHE~\cite{Yang_NatMater_2021}, and the SSE~\cite{Jungwirth_PRL_2021} (or MSHE~\cite{Kimata_Nature_2019, Hoffmann_PRL_2020}). The spin polarization of the spin current from the SHE is strictly aligned to the $y$-direction when we apply the electric field along to $x$-direction. Whereas for AFM-SHE~\cite{Yang_NatMater_2021}, the spin polarization, for example the $z$-polarized component, can be defined by the N\'{e}el vector $\mathbf{n}$ and spin-orbit field $\mathbf{H_{SO}}$, due to local inversion symmetry breaking in the case of Mn$_{2}$Au, as $\bm{\sigma_{z}}\propto\mathbf{n}\times\mathbf{H_{SO}}\perp\mathbf{n}$. On the other hand, the polarization of the spin current generated via the SSE is parallel to the N\'{e}el vector, i.e., $\bm{\sigma_{z}}\parallel\mathbf{n}$. In this study, we focus on collinear antiferromagnetic RuO$_{2}$, which theoretically is expected to have the SSE~\cite{Jungwirth_PRL_2021}, in order to reveal the actual physical mechanism. We systematically extract all $x$-, $y$-, and $z$-polarized components of the DL torque in RuO$_{2}$(100), (101), (001). We find the anisotropic DL torque in all of the components and the $z$-polarized DL

\begin{figure}[H]
	\includegraphics[width=0.5\textwidth]{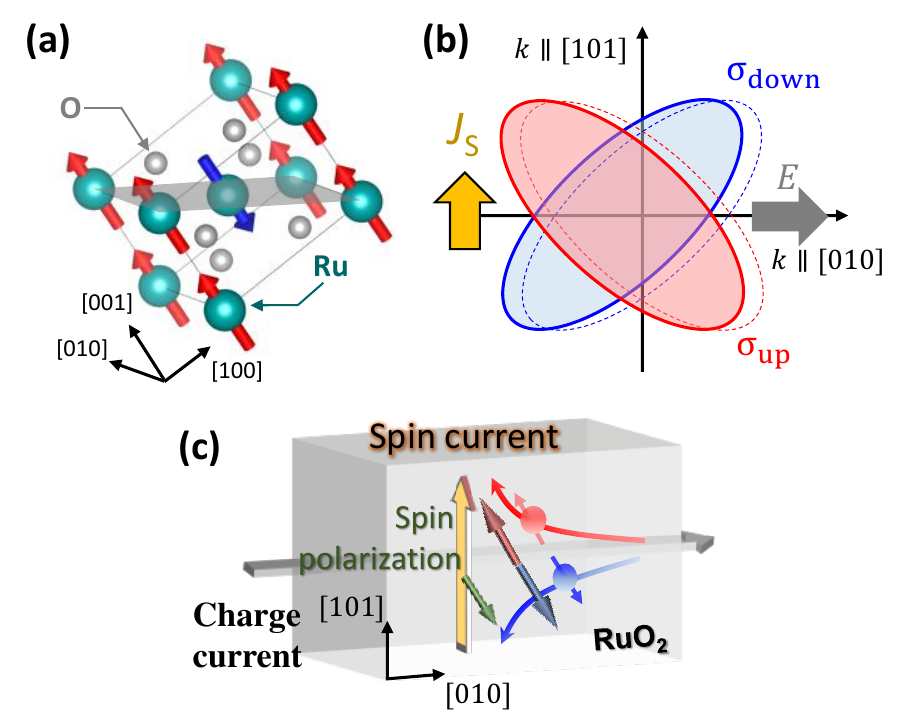}
	\caption{(a) Crystal structure of RuO$_{2}$ with the magnetic moment mostly aligned to [001] or [00$\bar{1}$] on the A and B sites of the Ru atoms. The specific crystal plane using a gray color, corresponds to (101). (b)Spin-split band originating from the RuO$_{2}$ antiferromagnetic order at the Fermi level. When we apply the electric field $\mathit{E}$ along the [010] or [0$\bar{1}$0] direction, the spin current $\mathit{J}_\mathrm{S}$ flowing in the [101] direction can be generated via the spin-splitter effect. (c) Schematic image of spin current generation in RuO$_{2}$(101). }
	\label{fig:1}
\end{figure}

\noindent torques, in particular RuO$_{2}$(101) when applying the charge current along the [010] direction. Further, using the $\mathit{z}$-polarized DL torque, we demonstrate field-free magnetization switching in adjacent magnetic layer.

Ruthenium dioxide (RuO$_{2}$) has recently been found to be a collinear antiferromagnet through superexchange coupling between Ru and O ions by means of neutron and X-ray diffraction profiles, which has a rutile-type crystal structure, as shown in Fig. 1(a) (space group: $\mathit{P}4_2/mnm$)~\cite{Weitering_PRL_2017, Comin_PRL_2019}. The N\'{e}el temperature was over 300 K~\cite{Weitering_PRL_2017}. The ruthenium oxide is electrically-conductive, which originates from the spin density wave instability at the Fermi surface~\cite{Weitering_PRL_2017}. The measured conductivity in this study is very similar to those of metals [see Fig. S2 in the Supplemental Material (SM)]. Further, several notable properties regarding the Dirac nodal line (DNL) electronic topology~\cite{Moser_PRB_2018}, strain-induced superconductivity~\cite{Uchida_PRL_2020} near 1 K, and crystal Hall effect~\cite{Sinova_SciAdv_2020, Jungwirth_arXiv_2021} in strong magnetic fields have been discovered recently. According to the reported diffraction measurements~\cite{Weitering_PRL_2017, Comin_PRL_2019}, the N\'{e}el vector is along [001] or [00$\bar{1}$], but is not strictly aligned and somewhat canted from this direction. Regarding the magnetic order, the spin current, with the spin polarization aligned to the N\'{e}el vector, is driven using a spin-split band structure, as shown in Fig. 1(b). When we apply an electric field to a specific direction, in this case [010], spin current flows along [101], with not only $\mathit{y}$-polarization like conventional SHE, but also $\mathit{x}$- and $\mathit{z}$-polarization, as shown in Fig. 1(c). This is advantageous for practical magnetization switching without the assistance of a magnetic field. 

\begin{figure}
	\includegraphics[width=0.5\textwidth]{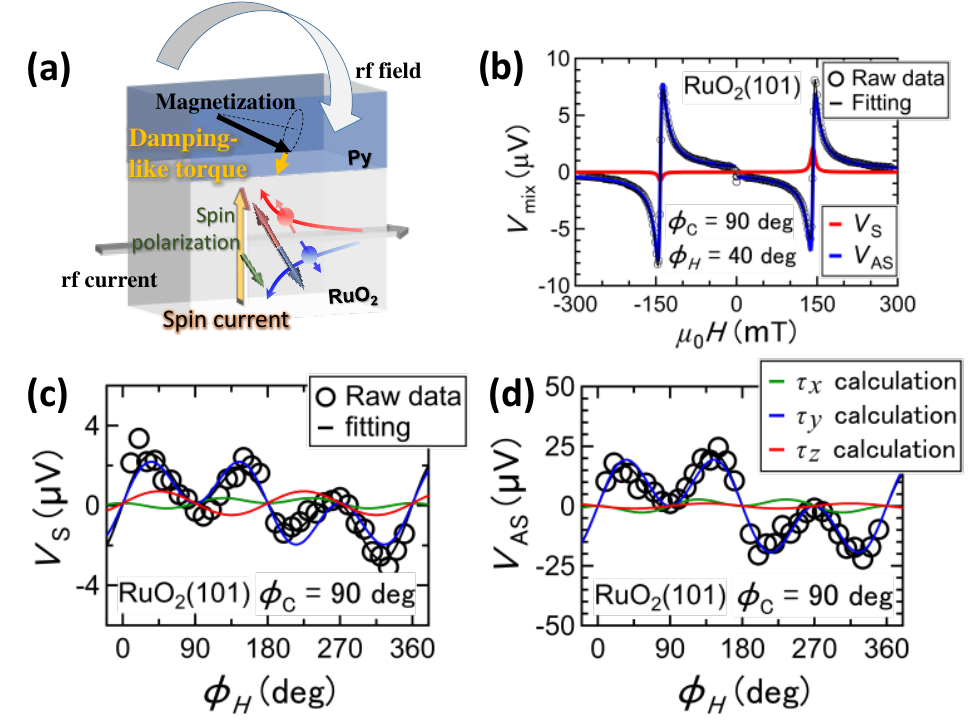}
	\caption{(a) Schematic image of RuO$_{2}$(101)/Py bilayer structure during the ST-FMR measurement. In the resonance field, the magnetization uniformly oscillates due to the rf Oersted field from the RuO$_{2}$ layer, while the spin current is generated via the SSE and injected into the Py layer. (b) ST-FMR raw data as a function of applied in-plane magnetic field. Black plots show the raw data. Black, red, and blue curves indicate fitting and calculations using the analyzed parameters. (c)Symmetric and (d)Antisymmetric voltage amplitudes depending on applied field angle for RuO$_{2}$(101) in the case where $\phi_{\mathrm{C}}$ = 90 deg. Black plots indicate the extracted values from Fig. 2(b). Black, red, blue, and green curves represent the fitting and calculations of $\tau_x$, $\tau_y$, and $\tau_z$, respectively, from the analyses using Eqs. (1) and (2).}
	\label{fig:2}
\end{figure}

To elucidate the relationship between the N\'{e}el vector and the spin polarization of the generated spin current, we prepared 10 nm-thick RuO$_{2}$(100), (101), and (001), which have an in-plane, canted, and perpendicular aligned N\'{e}el vector, onto Al$_{2}$O$_{3}$(0001), Al$_{2}$O$_{3}$(1$\bar{1}$02), and TiO$_{2}$(001) single-crystalline substrates respectively, by means of rf magnetron sputtering in an Ar atmosphere (0.24 Pa), introducing partial O$_2$ pressure (0.06 Pa). Further the substrate heating was conducted at 400 \(^\circ\)C during the deposition. Using X-Ray Diffraction (XRD) and Reflection High Energy Electron Diffraction (RHEED), we evaluated the crystallinity of the deposited films and succeeded in preparing epitaxial films of all the RuO$_{2}$ [see Fig. S1 in the SM]. The longitudinal resistivities $\rho_{xx}$ of the deposited RuO$_2$(100), (101), and (001) are approximately $180\: \mathrm{\mu\Omega cm}$, $60\: \mathrm{\mu\Omega cm}$, and $55\: \mathrm{\mu\Omega cm}$, respectively [see Fig. S2 in the SM]. Further, we deposited 5 nm-thick Ni$_{80}$Fe$_{20}$ (Py), as a spin detector, and 2 nm-thick AlO$_x$, as a capping layer in-situ. 
Based on the films, we fabricated rf waveguide devices by means of lift-off process using photolithography and Ar ion milling. First, we employed spin-torque ferromagnetic resonance (ST-FMR)~\cite{Liu_PRL_2011, Kurebayashi_NatNanotechnol_2011} at room temperature (RT), while applying an rf current at 2 dBm and 10 GHz, as shown in Fig. 2(a), to extract the DL torque with $\mathit{x}$-, $\mathit{y}$- and $\mathit{z}$-polarizations, because this is one of the most reliable techniques for decomposing the three terms~\cite{Tsymbal_NatComm_2020, Ralph_NatPhys_2017}. Figure 2(b) shows raw data of the detected ST-FMR signal $V_{\mathrm{mix}}$ as a function of the applied in-plane magnetic field in the case of RuO$_{2}$(101). Here, $\phi_H$ in Fig. 2(b) corresponds to the angle between the field and the applied current direction.  $\phi_{\mathrm{C}}$ is the angle between the applied current direction and the specific crystal direction ([010], [$\bar{1}$01], and [010] for RuO$_{2}$(100), (101), and (001), respectively). The detected signal basically consists of the Lorentzian $L$ and its derivative,  $V_{\mathrm{mix}} = V_{\mathrm{S}}L(H)+V_{\mathrm{AS}}\partial_HL(H)$~\cite{Liu_PRL_2011}, where $V_{\mathrm{S}}$ and $V_{\mathrm{AS}}$ correspond to the amplitudes of the Lorentzian and its derivative, respectively. As shown in Fig. 2(b), the fitting was successfully and decomposed into both the Lorentzian and its derivative. Based on the analysis, we summarized the amplitudes as a function of $\phi_{H}$, as shown in Figs. 2(c) and (d). The $x$-, $y$- and $z$-components can be separated from the $\phi_{H}$ dependence using Eqs. (1) and (2)~\cite{Tsymbal_NatComm_2020, Ralph_NatPhys_2017}.
\begin{equation}\label{eq:1}
\begin{split}
    V_{\mathrm{S}}(\phi_H)
    &\propto \sin 2\phi_H \left[\tau_{x}^{\mathrm{DL}}\sin \phi_H +\tau_{y}^{\mathrm{DL}}\cos \phi_H+\tau_{z}^{\mathrm{FL}}\right],
\end{split}
\end{equation}
\begin{equation}\label{eq:2}
\begin{split}
    V_{\mathrm{AS}}(\phi_H)
    &\propto\sin 2\phi_H \left[\tau_{x}^{\mathrm{FL}}\sin \phi_H +\tau_{y}^{\mathrm{FL}}\cos \phi_H+\tau_{z}^{\mathrm{DL}}\right]
\end{split}
\end{equation}
Here $\tau_i^{\mathrm{DL}}$ and $\tau_i^{\mathrm{FL}}$ represent the amplitudes of the DL and field-like (FL) torque, respectively, where subscript $i$ represents the $x$, $y$, and $z$ components. Through the analyses using Eqs.(1) and (2), the amplitudes were clearly fitted and decomposed into each $\tau_i^{\mathrm{DL}}$ or $\tau_i^{\mathrm{FL}}$. The existences of a NON $\mathit{y}$-polarization component is important because this cannot be with the conventional SHE due to the strict orthogonal relationship between the applied charge current, generated spin current, and spin polarization~\cite{Dyakonov_JETP_1971, Kimura_PRL_2007, Saitoh_APL_2006}. Therefore, this implies that the ST-FMR signal includes both SSE and SHE. 

Based on the extracted amplitudes $V_{\mathrm{S}}$ and $V_{\mathrm{AS}}$, we estimated the DL torque efficiency per unit electric field using Eqs. (3) and (4)~\cite{Liu_PRL_2011}.
\begin{equation}\label{eq:3}
\begin{split}
    \xi_{i}^{\mathrm{DL,}E}
    &=\frac{\tau_i^{\mathrm{DL}}}{\tau_y^{\mathrm{FL}}}\frac{e\mu_0M_{\mathrm{S}}t_{\mathrm{Py}}t_{\mathrm{RuO_2}}}{\hbar\rho_{xx}}\sqrt{1+\frac{\mu_0M_{\mathrm{eff}}}{\mu_0H_{\mathrm{R}}}},
\end{split}
\end{equation}
\begin{equation}\label{eq:4}
\begin{split}
    \xi_{z}^{\mathrm{DL,}E}
    &=\frac{\tau_z^{\mathrm{DL}}}{\tau_y^{\mathrm{FL}}}\frac{e\mu_0M_{\mathrm{S}}t_{\mathrm{Py}}t_{\mathrm{RuO_2}}}{\hbar\rho_{xx}}
\end{split}
\end{equation}
Here $e$, $\mu_0M_{\mathrm{S}}$, $t_{\mathrm{Py}}$, $t_{\mathrm{RuO_2}}$, $\hbar$, $\rho_{xx}$, $\mu_0M_{\mathrm{eff}}$, and $\mu_0H_{\mathrm{R}}$ are the elementary charge, saturation magnetization of Py layer, thickness of Py, thickness of RuO$_{2}$ layer, the Dirac constant, longitudinal resistivity of RuO$_{2}$, effective magnetization of Py, and Py resonance field at 10 GHz, respectively. Note that $\sqrt{1+(\mu_0M_{\mathrm{eff}}/\mu_0H_{\mathrm{R}})}$ is not needed for $\xi_z^{\mathrm{DL,}E}$ in Eq. (4) because both $\tau_z^{\mathrm{DL}}$ and $\tau_y^{\mathrm{FL}}$ are from the same antisymmetric voltage $V_{\mathrm{AS}}$. The effective magnetization $\mu_0M_{\mathrm{eff}}$ (0.8 T) was extracted from the Kittel relation in this ST-FMR measurement. The saturation magnetization of the Py layer, $\mu_0M_{\mathrm{S}} = 1.1 $ T, prepared by us, was extracted from our previous report~\cite{Karube_JMMM_2020}.

\begin{figure}[b]
	\includegraphics[width=0.4\textwidth]{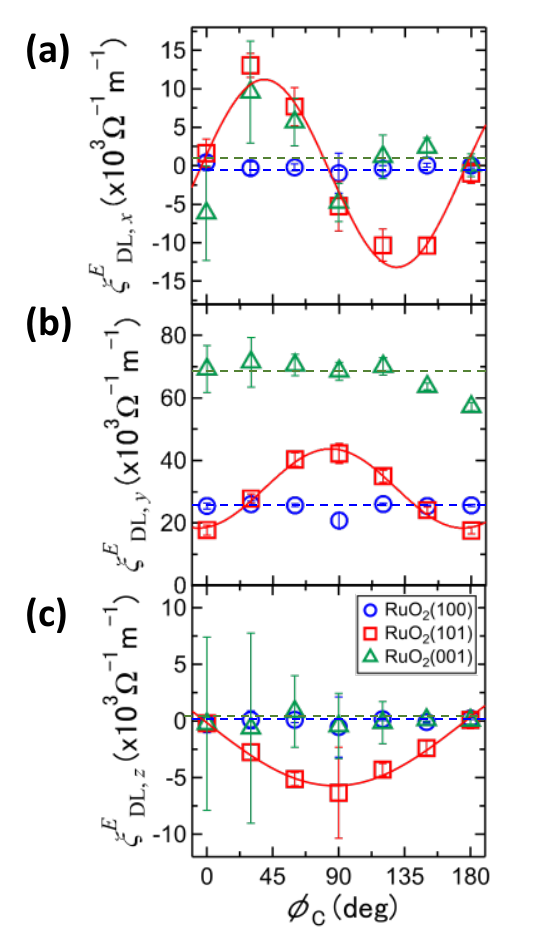}
	\caption{Damping-like torque efficiency per unit electric field as a function of crystal angle $\phi_{\mathrm{C}}$ for (a)$\mathit{x}$-, (b)$\mathit{y}$-, and (c)$\mathit{z}$-components. Blue open circle, red open rectangle, and green open triangle correspond to RuO$_{2}$(100), (101), and (001) cases, respectively.}
	\label{fig:3}
\end{figure}

\begingroup
\begin{table*}[!]
 \caption{Damping-like torque efficiencies from the spin Hall and spin-splitter effects in  RuO$_2$(101)}
 \label{table:SpeedOfLight}
 \centering
 \renewcommand{\arraystretch}{1.5}
  \begin{tabular}{c|cccc}
   \hline
    $i$ & SSE,\;$x$ & SSE,\;$y$ & SSE,\;$z$ & SHE,\;$y$ \\
   \hline
    $\xi_{i}^E$ [$\frac{\hbar}{2e}\:\Omega^{-1}\mathrm{m}^{-1}]$ & $(1.2\pm0.1)\times10^4$ & $(2.5\pm0.1)\times10^4$ & $-(5.8\pm0.5)\times10^3$ & $(1.8\pm0.8)\times10^4$ \\
    $\xi_{i}$ [$-$] & $0.0066\pm0.0005$ & $0.0138\pm0.0001$ & $-0.0032\pm0.0003$ & $0.0099\pm0.0044$ \\
    $\xi_{i}/\xi_{\mathrm{SHE,}y}$ [$-$] & $0.67$ & $1.39$ & $-0.32$ & $-$ \\
    \hline
  \end{tabular}
\end{table*}
\endgroup

The DL torque efficiencies for each component depending on the crystal angle $\phi_{\mathrm{C}}$ are shown in Fig. 3. These graphs show the relationship of the generated DL torque efficiency between the applied charge current and the N\'{e}el vector orientation, as defined above. Regarding RuO$_2$(100) and (001), the $x$-, and $z$-components generated from the SSE are negligible, as shown in Figs. 3 (a), and (c). On the other hand, a finite $y$-component of the DL torque efficiency was found and was nearly independent of the crystal angle $\phi_{\mathrm{C}}$, i.e. constant the $\xi_{\mathrm{DL,}y}^{E}$ for RuO$_2$(100) and (001) are $(2.5\pm0.2)\times10^4 \; [(\hbar/2e)\:\Omega^{-1}\mathrm{m}^{-1}]$ and $(6.7\pm0.5)\times10^4 \; [(\hbar/2e)\:\Omega^{-1}\mathrm{m}^{-1}]$, respectively (or $0.046\pm0.003, 0.038\pm0.002$ for the dimensionless efficiency $\xi_{\mathrm{DL,}y}$). Note that there is a difference in the resistivity between RuO$_{2}$ (100) and (001) [see the SM], causing a change in the amplitude relation between $\xi_{\mathrm{DL,}y}^{E}$ and $\xi_{\mathrm{DL,}y}$. Remarkably all components of the DL torque efficiency are finite and anisotropic on the crystal angle $\phi_{\mathrm{C}}$ in the case of RuO$_2$(101). The $x$-, $y$-, and $z$-components  follow Eqs. (5)-(7), respectively.
\begin{equation}\label{eq:5}
\begin{split}
    \xi_{\mathrm{DL,}x}^{E}(\phi_{\mathrm{C}}) = \xi_{\mathrm{SSE,}x}^E\sin\phi_{\mathrm{C}}\cos\phi_{\mathrm{C}}
\end{split}
\end{equation}
\begin{equation}\label{eq:6}
\begin{split}
    \xi_{\mathrm{DL,}y}^{E}(\phi_{\mathrm{C}}) = \xi_{\mathrm{SHE,}y}^E+\xi_{\mathrm{SSE,}y}^E\sin^2\phi_{\mathrm{C}}
\end{split}
\end{equation}
\begin{equation}\label{eq:7}
\begin{split}
    \xi_{\mathrm{DL,}z}^{E}(\phi_{\mathrm{C}}) = \xi_{\mathrm{SSE,}z}^E\sin\phi_{\mathrm{C}}
\end{split}
\end{equation}
Here $\xi_{\mathrm{SSE,}i}^E (i=x, y, z)$ and $\xi_{\mathrm{SHE,}y}^E$ are the amplitudes of the SSE and SHE for each component, respectively. We further investigated a 3 nm-thick Cu insertion case between the RuO$_2$ and Py layers to exclude the possibility of an exchange bias. We confirmed the same behaviors compared with the case without Cu insertion [see Fig. S3 in the SM]. 

The anisotropic and independent behaviors regarding the crystal angle $\phi_{\mathrm{C}}$ for all RuO$_2$(100), (101), and (001) could be explained by the orthogonal relation between the applied charge current, spin current, and spin polarization directions. In the SSE, the spin polarization of the generated spin current follows the N\'{e}el vector direction approximately~\cite{Weitering_PRL_2017, Comin_PRL_2019}. As for RuO$_2$(001) with a perpendicular N\'{e}el vector against the film plane, the spin polarization and flowing spin current directions are parallel to each other, thus the spin current can no longer occur. Therefore, we found that only the $y$-component of the DL torque generated by the SHE, as shown in Fig.3 (b). Further, in the case of RuO$_2$(101) with a canted N\'{e}el vector, all components of the torque may appear and should be anisotropic to the N\'{e}el vector direction. The possible contribution of spin-orbit precession (SOP) torque~\cite{Hibino_APLMater_2020} generated at the RuO$_2$/Py interface has been considered. This torque may have non $y$-components of the DL torque due to precessional motion being induced by the Rashba effect at the interface. Note that now the effective field generated by the Rashba effect is always perpendicular to the applied current direction; the SOP should be independent of $\phi_{\mathrm{C}}$, therefore, we conclude that the SOP effect is negligible in this RuO$_2$/Py bilayer. Regarding RuO$_2$(100) with an in-plane N\'{e}el vector, the $z$-component should not appear, while the $x$-component may be generated when the applied current and the N\'{e}el vector have relative angles on the in-plane; however, as shown in Fig.3 (a), this was not observed. The possible reason for negligible $x$-component spin current in RuO$_2$(100) is due to higher scattering rate $\Gamma$ than that of RuO$_2$(101). The SSE is related to the $\mathscr{T}$-odd spin Hall conductivity depending on the $\Gamma$. The estimated $\Gamma$ from $\rho_{xx}$ in RuO$_2$(100) is 31 meV, while $\Gamma$ in RuO$_2$(101) is 9.5 meV. The quantitative comparison with the theory~\cite{Jungwirth_PRL_2021} is still difficult, but it predicts that the scattering rate suppresses the SSE. Next, we estimated all amplitudes for RuO$_2$(101), as shown in TABLE I, based on the angular dependence. The amplitude ratio between the SSE and SHE is larger or comparable for all $x$-, $y$-, and $z$-components. Thus, we have efficiently found finite but small $x$-, $y$-, and $z$-components of the DL torque, even though the theory for the SSE predicted much higher efficiency values, approximately $30\; \%$~\cite{Jungwirth_PRL_2021}. This result is attributed to the AFM domain structure~\cite{Radaelli_NatMater_2018} cancelling out the unidirectional N\'{e}el vector contribution. Although we have found 2-fold in-plane crystal symmetry in RuO$_2$(101) [see Fig. S4 in the SM], the two peaks are broad, implying that the prepared RuO$_2$ is distorted. This could lead to the AFM domain structure reducing the unconventional torque originating from the SSE, because the distortion would make the AFM locally ordered due to weak superexchange coupling between the Ru and O ions. Further, for the amplitude comparison of the detected $y$-components among RuO$_2$(100), (101), and (001), we are currently unable to fully understand the relation even when consider the contribution of the expected DNL topology~\cite{Moser_PRB_2018}. When the electrons in RuO$_2$ travel on the (110) and ($\bar{1}$10) planes, the Berry curvature driven by the DNL topology occurs, enhancing the $y$-component of the DL torque efficiency. In this case, we should see the enhancement of the torque near the (110) and ($\bar{1}$10) planes, dependent on the crystal angle $\phi_{\mathrm{C}}$. 

\begin{figure*}
	\includegraphics[width=1.0\textwidth]{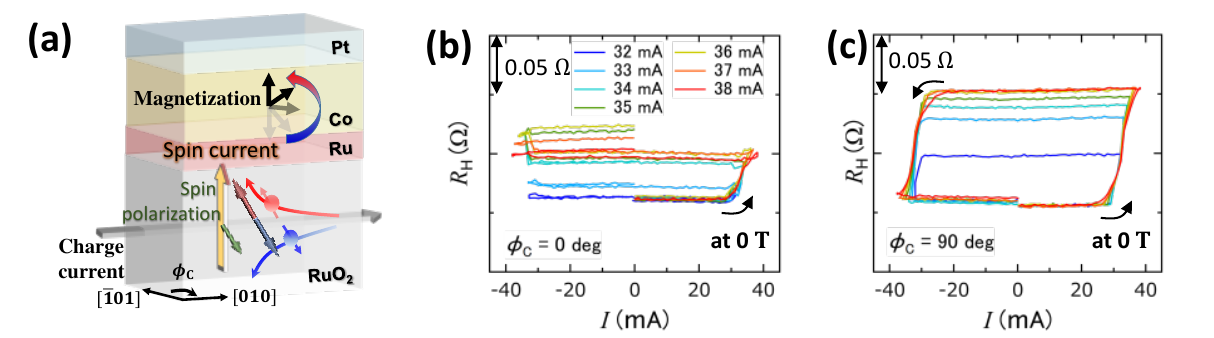}
	\caption{(a)Schematic illustration of RuO$_2$(101)(10nm)/Ru(0.8nm)/Co(0.8nm)/Pt(2nm) multilayer. The $z$-polarized spin current generated via the SSE is injected into the Co layer, then the magnetization is switched without the assisting external magnetic field. Hall resistance of the Co layer when applying the DC current $I$ for the switching in the case of (b) $\phi_{\mathrm{C}} =$ 0 deg and (c) 90 deg.}
	\label{fig:4}
\end{figure*}

Next we demonstrated the field-free switching using the $z$-polarization component of the DL torque in RuO$_2$(101) at RT. For the experiment, we fabricated a RuO$_2$(101)(10 nm)/Ru(0.8 nm)/Co(0.8 nm)/Pt(2 nm) multilayer structure, as shown in Fig. 4(a). The roles of the Ru and Pt layers are to break the RuO$_2$ crystallinity and to introduce a perpendicular magnetic anisotropy (PMA) of the Co layer through  hexagonal closed packed structure and spin-orbit coupling. Note that the 2 nm-thick Pt layer would not contribute to the switching of the Co layer through the generated spin-orbit torque because most of the charge current flows into the 10 nm-thick RuO$_2$(101) layer due to current shunting. The calculated shunting factors for the Pt and RuO$_2$(101) layers are 0.06 and 0.94 [see the SM], respectively, thus, the current dominantly flows into the RuO$_2$ layer. Furthermore, the spin current generation is not sufficient in the Pt layer due to the comparable thickness of the spin diffusion length ($\lambda_{\mathrm{sd}}^{\mathrm{Pt}} \approx 1.0 $\: nm)~\cite{Karube_PRApplied_2020}. As shown in Fig. 4(a), when we apply the charge current along [$\bar{1}$01] direction ($\phi_{\mathrm{C}} = 0$ deg), without the $z$-polarization DL torque, switching cannot fully occur even when applying large current (Fig. 4(b)). When the current is applied [010] direction ($\phi_{\mathrm{C}} = 90$ deg), with the $z$-polarization component, the full switching is observed (Fig. 4(c)); therefore, $z$-polarization is necessary for field-free switching. To ensure occurrence of the $z$-component, we confirmed an effective perpendicular field $\mu_0H_z^{\mathrm{eff}}$ depending on the current amplitude along [010] direction [see Fig. S5 in SM]. On the other hand, halfway switching, in the case of $\phi_{\mathrm{C}} = 0$ deg, as shown in Fig. 4(b), would originate from the perpendicular exchange bias. We have only observed the finite exchange bias where $\phi_{\mathrm{C}} = 0$ deg, whereas it was negligible when $\phi_{\mathrm{C}} = 90$ deg [see the SM as shown in Fig. S6]. This bias prevents reverse switching due to pinning the magnetization along the bias direction. A finite $y$-component of the DL torque turn the magnetization, but is unable to fully switch it without the $z$-component of the DL torque generated via the SSE. In the case where $\phi_{\mathrm{C}} = 90$ deg, we observed memristive behavior with a change in the current amount. The observed memristive behavior is explained in terms of multi-domain structure in the PMA Co layer induced by exchange coupling between AFM RuO$_2$ and ferromagnetic Co layers as is previously reported in a PtMn/[Co/Ni]$_n$ structure~\cite{Fukami_NatMater_2016}. We emphasize that the demonstration of the field-free switching does not originate from the exchange bias, but the $z$-component of the DL torque generated from RuO$_2$(101). The in-plane exchange bias is induced along the [$\bar{1}$01] direction ($\phi_{\mathrm{C}} = 0$ deg) due to the canted N\'{e}el vector. Hereby, we have successfully demonstrated the generation of spin-splitter torque, including all $x$-, $y$-, and $z$-components of the DL torque and the field-free switching driven by the unconventional torque.

In summary, we prepared collinear antiferromagnetic RuO$_2$(100), (101), and (001) grown epitaxially, where the SSE is theoretically expected to generate the unconventional DL torque~\cite{Jungwirth_PRL_2021}. Using the epitaxial films, all $x$-, $y$-, and $z$-components of the DL torque in the ruthenium oxides have been systematically investigated by means of the ST-FMR technique. Interestingly, the finite unconventional $x$- and $z$-components of the DL torque, depending on the N\'{e}el vector direction, have been observed in RuO$_2$(101). These components clearly originate from the SSE. Using the $z$-component of the DL torque from RuO$_2$, we have demonstrated field-free switching in the FM layer with the PMA. So, this study provides a novel spin current generation and a technique to manipulate the spins effectively for antiferromagnetic spin-orbitronics.



\begin{acknowledgments}
We would like to thank Prof. Akimasa Sakuma, Dr. Chaoliang Zhang, and Yuta Yahagi for constructive disccusions. Further, this work is partially supported by the Japan Society for the Promotion of Science (JSPS) (Grants No. 15H05699, and No. 18K14111), the Center for Spintronics Research Network at Tohoku University, and the Center for Science and Innovation in Spintronics at Tohoku University.
\end{acknowledgments}

\bibliographystyle{apsrev4-2}
\bibliography{refs}


\clearpage

\begin{figure*}
	\includegraphics[width=1.0\textwidth]{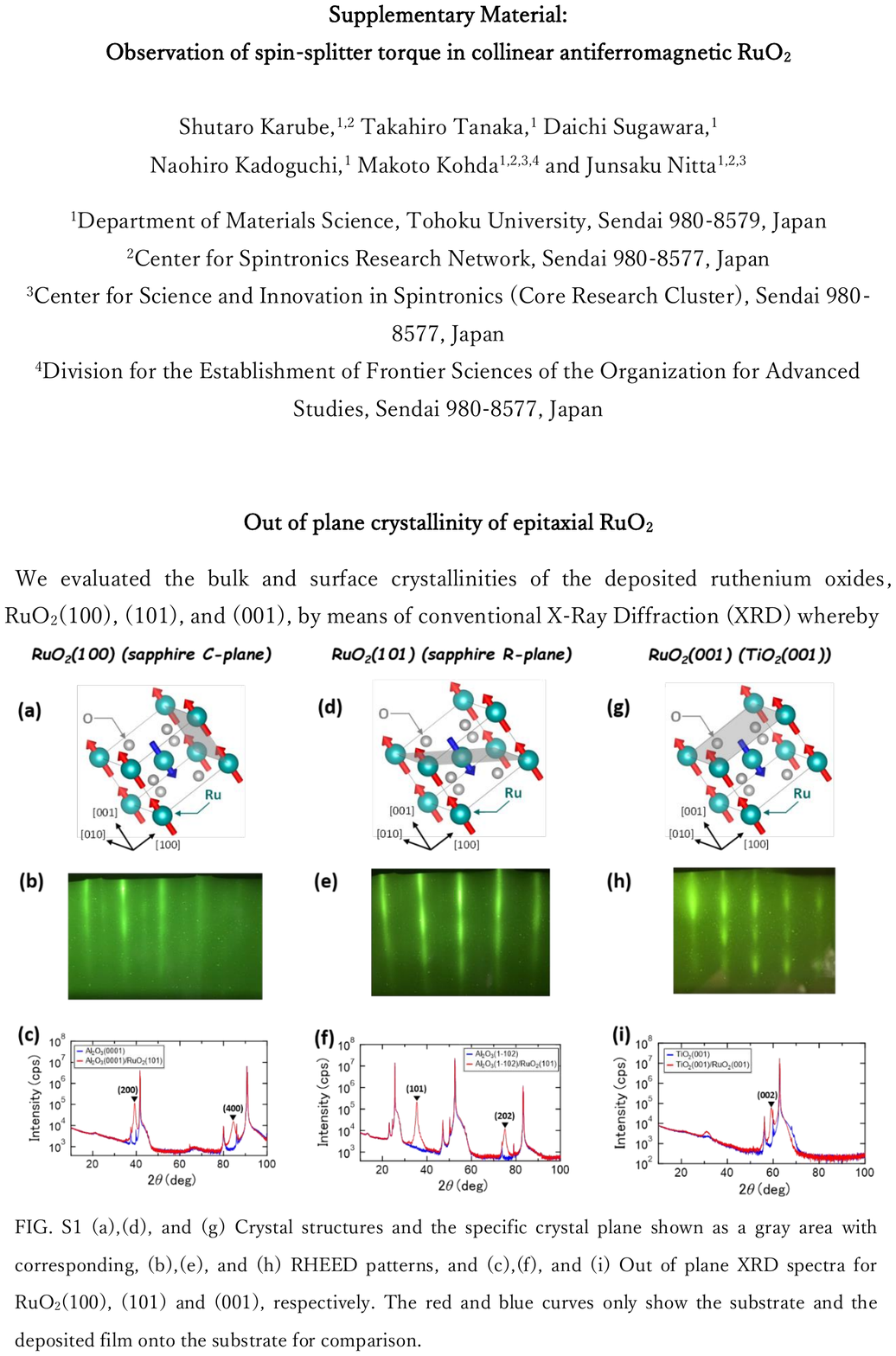}
\end{figure*}

\begin{figure*}
	\includegraphics[width=1.0\textwidth]{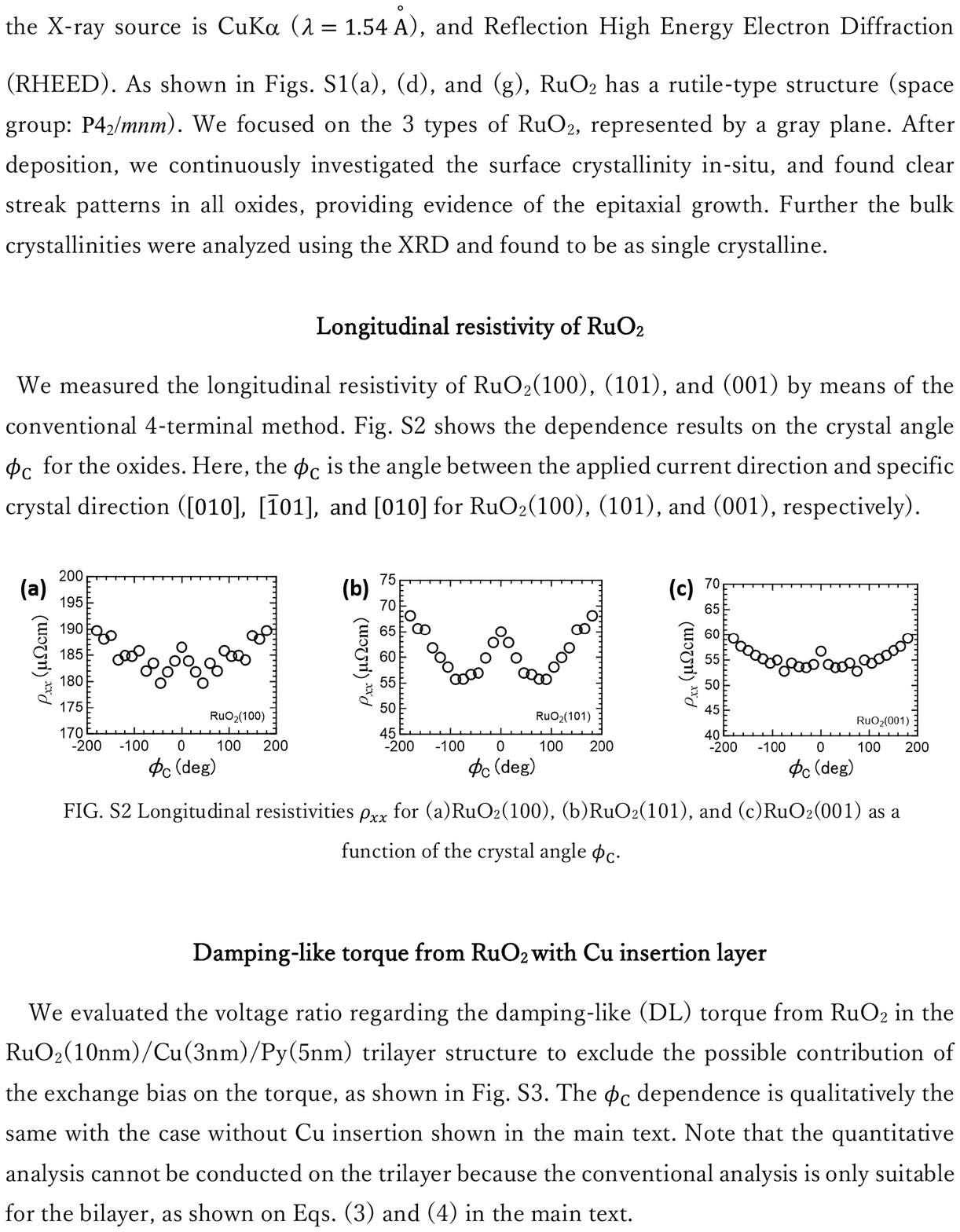}
\end{figure*}

\begin{figure*}
	\includegraphics[width=1.0\textwidth]{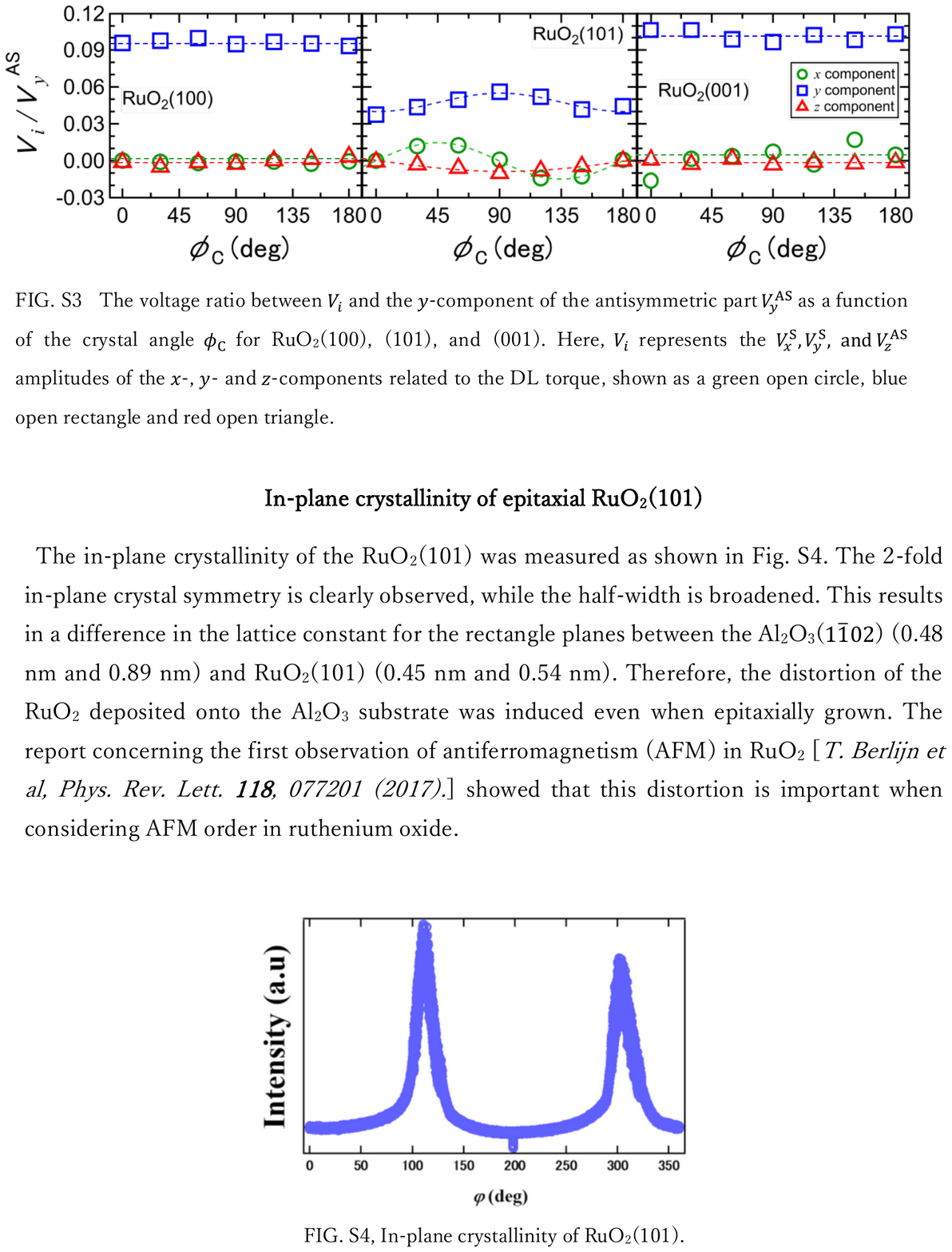}
\end{figure*}

\begin{figure*}
	\includegraphics[width=1.0\textwidth]{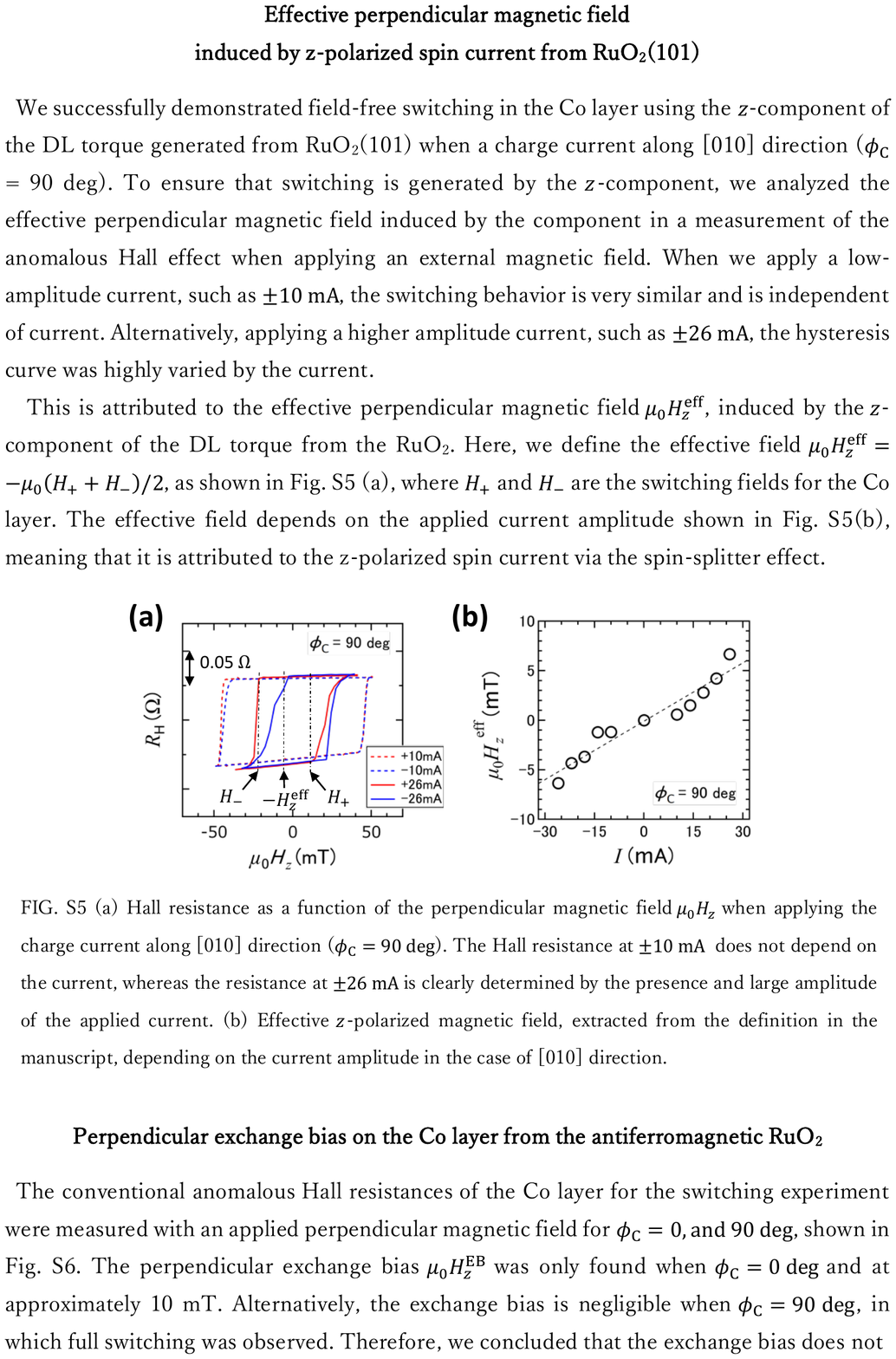}
\end{figure*}

\begin{figure*}
	\includegraphics[width=1.0\textwidth]{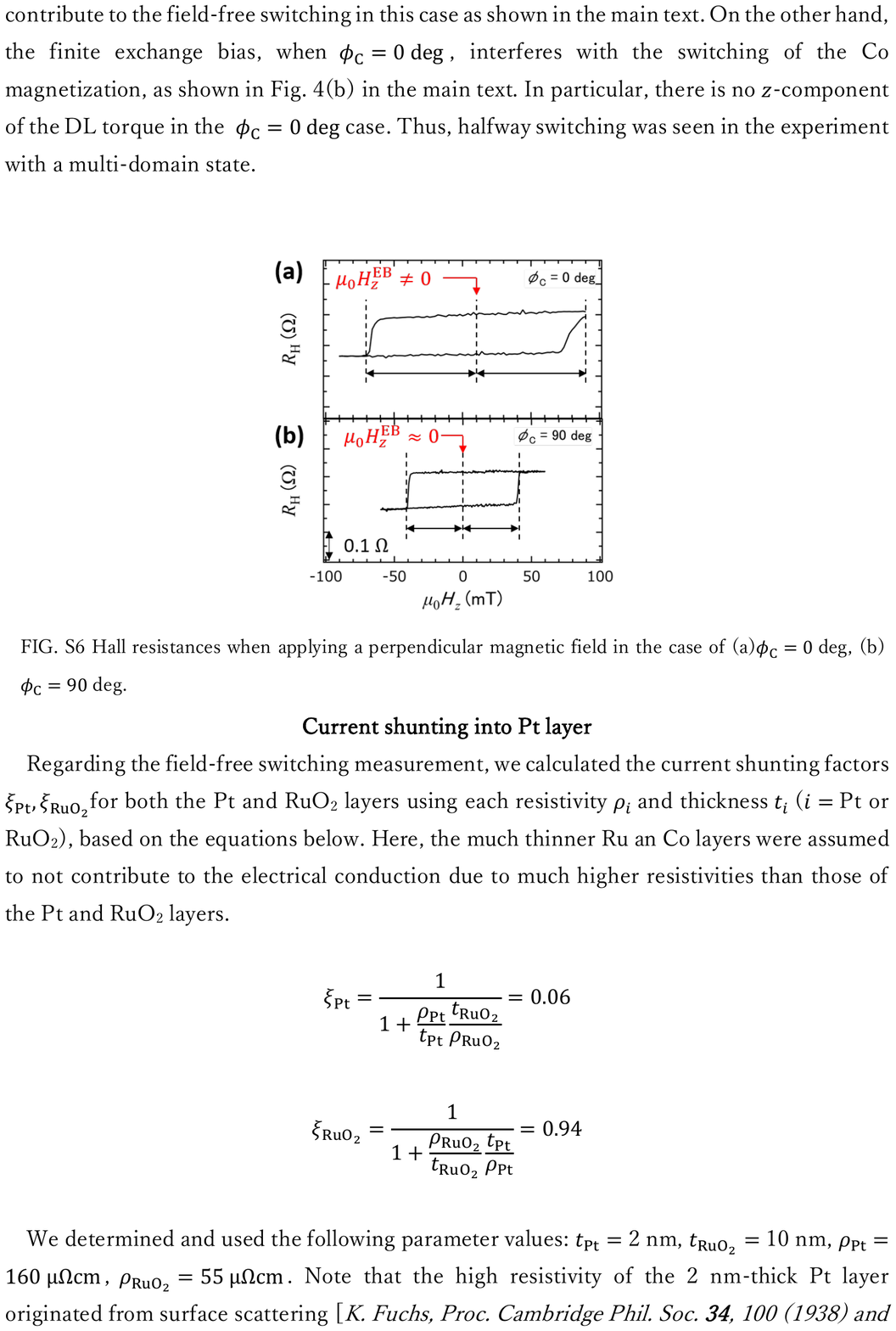}
\end{figure*}

\begin{figure*}
	\includegraphics[width=1.0\textwidth]{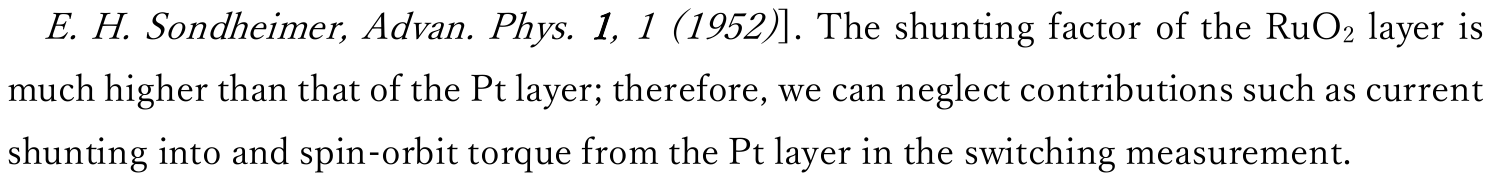}
\end{figure*}

\end{document}